\documentclass[aps,pra,twocolumn,groupedaddress,epsfig,amsmath,floatfix,showpacs]{revtex4}
\usepackage{epsfig,amsmath}
\usepackage{graphicx}
\usepackage{dcolumn}
\usepackage{bm}
\begin{document}


\title{Momentum Analysis in Strong-field Double Ionization}
\author{P. J. Ho\footnote{Electronic address: phayho@pas.rochester.edu}}
\affiliation{{Department of Physics and Astronomy,} \\
{University of Rochester, Rochester, NY 14627}}
\date{\today}

\begin{abstract}

We provide a basis for the laser intensity dependence of the momentum distributions of electrons and ions arising from strong-field  non-sequential double ionization (NSDI) at intensities in the range $I=1-6.5 \times 10^{14} W/cm^2$. To do this we use a completely classical method introduced previously \cite{ho-etal05}.  Our calculated results reproduce the features of experimental observations at different laser intensities and depend on just two distinct categories of electon trajectories.

\end{abstract}

\maketitle

 \section{Introduction} 

\begin{figure}
\centerline{\includegraphics[width=1.5in]{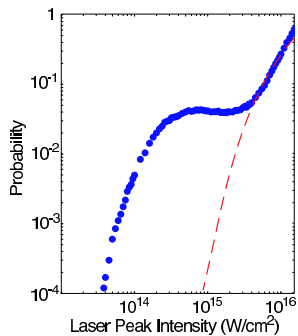}}
\caption{Points trace out the knee-like structure of doubly-ionized ion yield in strong-field double ionization.  The dashed curve (red) is inserted by hand to indicate the prediction made by sequential quantum tunneling theory \cite{PPT-ADK}.}
\label{fig.knee}
\end{figure}

When atoms with two or more electrons are exposed to an ultra-short (10-600 femtoseconds) and ultra-strong ($1-100 \times10^{14} W/cm^2$) laser pulse, two electrons can be ripped off from the nucleus.  This strong-field double ionization process has been observed in all inert gas atoms \cite{fittinghoff-etal, walker-etal,NSDI-atoms} and some molecules \cite{NSDI-molecules}.  The remarkable aspect of this process is its strong electron pair correlation.  Uncorrelated electrons are ionized sequentially, each event well explained by single-ionization theory \cite{PPT-ADK}.  This sequential ionization theory cannot explain a characteristic knee-like structure observed in the double-ionization yield (see Fig. \ref{fig.knee}) as a function of laser intensity, so this process is usually known as nonsequential double ionization (NSDI).  

\begin{figure}
\centerline{\includegraphics[width=3.2in]{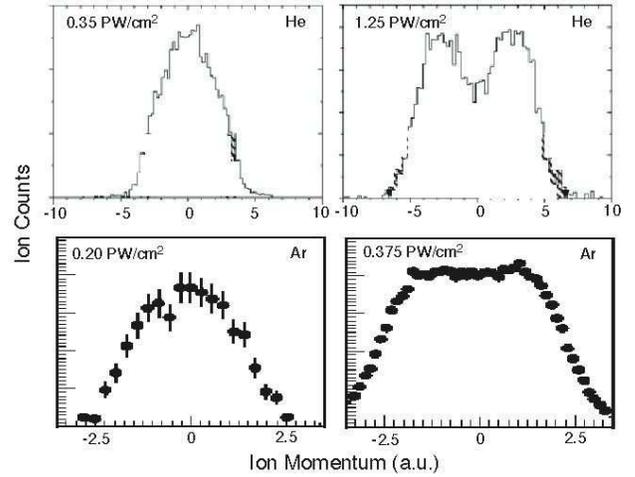}}
\caption{Experimental ion momentum distributions along the longitudinal direction for He$^{2+}$ \cite{Ullrich-04} and Ar$^{2+}$ \cite{weber_Ar} at different laser intensities.  As the laser intensity increases, the width of the distribution become broader and may develop into a double peak.  We will comment on the experimental ion momentum distributions of Ne$^{2+}$ at the end of this paper.}
\label{fig.pdist-expt}
\end{figure}

In addition to ion yield, experiments in the intensity regime of the knee-like structure have examined recoil ion momentum distributions in which the shape depends strongly on laser intensity \cite{weber_He, weber_Ar, feuerstein_Ar, eremina, Ullrich-jpb04, Ullrich-04, moshammer_Ne}.  Distributions for different atomic species (helium, neon and argon) show different shape variations as a function of laser intensity, as shown in Fig. \ref{fig.pdist-expt}: at low laser intensities there are more low-momentum or zero-momentum ions, giving a narrow distribution peaked at zero momentum, whereas at higher laser intensity the high-momentum ions become more dominant and the distribution becomes broader.

\begin{figure*}
\centerline{\includegraphics[width=6in]{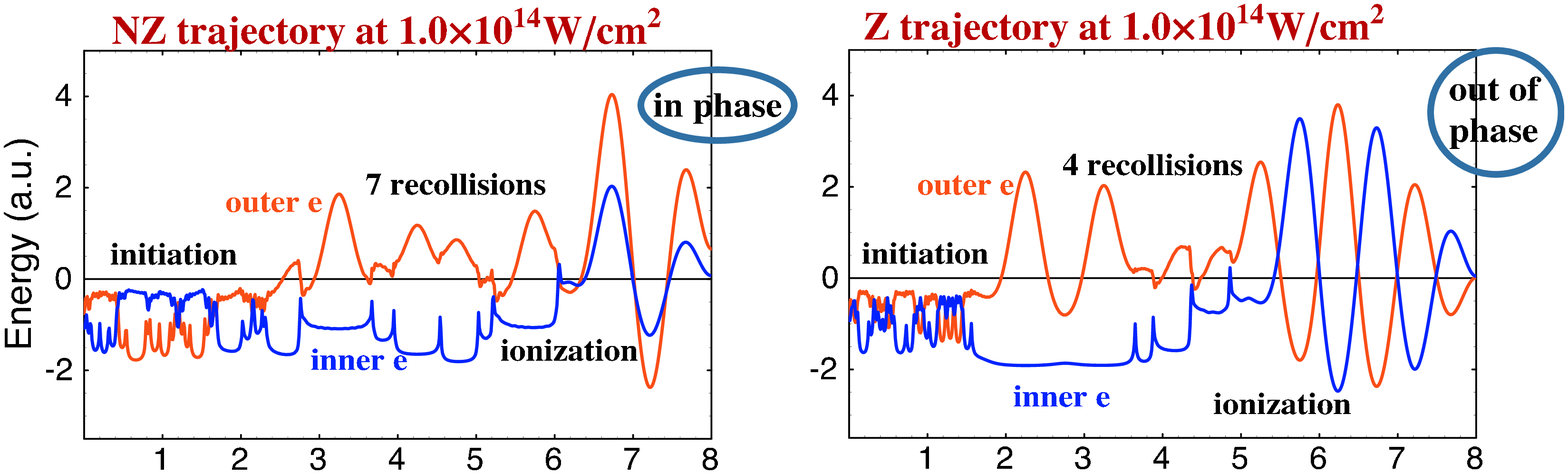}}
\caption{Plots of electron energy versus time measured in laser cycles, showing four distinct stages of NSDI. We show a variety of histories of two-electron trajectory that lead to double ionization.  One can identify four universal stages, which are embedded in these different histories of evolution.  A trajectory is labelled as NZ and Z if both electrons moving away from the nucleus in the same and opposite directions respectively.  See \cite{ho-etal05} for detailed decription of the four stages.}
\label{fig.energyZNZ}
\end{figure*}

A variety of theoretical initiatives have been pursued  \cite{corkum, taylor, smatrix, coulomb-focusing, softcollision, Texas} with the goal of understanding NSDI effects including ion yield and final-state momenta.  With the exception of the calculations made by the Taylor group that are applicable only to helium \cite{taylor}, all have {\em ad hoc} elements that are not easy to justify fully, and none have successfully dealt with the intensity-dependence of the shapes of the experimental momentum distributions shown in Fig. \ref{fig.pdist-expt}.

Recently we pursued a fully classical theory \cite{ho-etal05} as an explanation for the electron correlation in NSDI and suggested that the major features of NSDI arise naturally without needing to engage quantum processes at all, not even tunnelling as the initial step in NSDI.  In such a treatment the NSDI process is a consequence of both electrons actively responding to the laser field and to each other and to the nuclear attraction, undergoing ordinary (but complicated) Hamiltonian dynamics throughout the duration of the laser pulse. As one consequence, double ionization can be easily seen to occur in four universal stages (see Fig. \ref{fig.energyZNZ}).

In this brief note we provide calculational evidence supporting a classical interpretation of the intensity-dependent changes in the shape of ion recoil momentum distributions that are shown in Fig. \ref{fig.pdist-expt}. The calculations suggest the active presence of two distinct categories of two-electron ejection from the atom, loosely speaking associated with back to back ejection on the one hand and side by side ejection on the other hand. In the first case one can expect that little net momentum will be carried by the pair and in the second case the net electron momentum can be substantial. For ease of description we have labelled these as the Z and NZ categories \cite{Ho-Eberly03,ho-etal05}, meaning ``zero" and ``non-zero" momentum to impart to the ion.  

Our approach will be to display the results of entirely classical calculations of the NSDI electron trajectories as evidence that the Z-NZ distinction exists, and then show that it explains the origin of the laser intensity dependence of the shape of the momentum distributions shown in Fig. \ref{fig.pdist-expt}.

 \section{Classical Ensemble Method} 

We use the classical ensemble method that has been described in detail by Panfili, {\it et al.}  \cite{Panfili-etal01}.  The idea is to use a large number of classical electron pairs (100,000 - 500,000 members) to model the dynamical evolution of a two-electron system exposed to an intense laser field.  

We are interested only in the forward-backward aspects of the recoil ion momentum, what is measured as the longitudinal momentum (i.e., along the axis of the laser polarization). This means that  the one-dimensional aligned-electron approximation (AEA) \cite{Javanainen-etal87}, in which the response of each electron is restricted to the axis of the laser polarization, will be adequate. All of the relevant experimental data is for linearly polarized laser light, which is the vast majority of available data. We also use the so-called Rochester screened Coulomb potential \cite{Javanainen-etal87} $V(x) = -1/\sqrt{x^2 + 1}$ to model the nuclear binding potential and electron-electron repulsion potential.  We have previously shown that the character of the NSDI process (recall Fig. \ref{fig.energyZNZ}) obtained using this 1d model does not differ much from that of 3d \cite{ho-etal05}.  In addition, Haan, {\it et al.} have shown that there are striking similarities in the dynamical evolution of the classical ensemble calculations and the corresponding quantum 2-e wave functions \cite{Panfili-etal02, haan-etal02}. 

Each electron pair is given a different set of initial conditions but each pair is given an initial energy of -2.24 a.u., which corresponds to the energy of the two-electron quantum ground state. We use $E(t) = {\cal{E}}_0f(t)\sin\omega t$ to denote the optical field, where ${\cal{E}}_0$ is the maximum field strength, $f(t)$ is the envelope function. We take an 8-cycle ($\sim$ 25 fs) sinusoidal laser pulse with the wavelength 780 nm (frequency $\omega$ = 0.0584 a.u.) and a trapezoidal envelope with 2-cycle turn-on and turn-off.  

The response of the two electrons can be obtained by solving two coupled Newtonian equations of motion: 
\begin{eqnarray} 
\ddot x_1&=&-E(t)-\frac{2x_1}{(x_1^2+1)^{3/2}} +\frac{(x_1-x_2)}{((x_1-x_2)^2+1)^{3/2}},\label{eqn:ode1}\nonumber \\ 
\ddot x_2 & =& -E(t)-\frac{2x_2}{(x_2^2+1)^{3/2}} -\frac{(x_1-x_2)}{((x_1-x_2)^2+1)^{3/2}} \nonumber
\label{eqn:ode2} . 
\end{eqnarray} 
Here $x_1$ and $x_2$ are the positions of two electrons, and clearly the evolution of each electron pair can be visualized as a distinct trajectory in the $x_1-x_2$ plane.  In this treatment all Coulombic forces and laser-electron forces are acting continuously and simultaneously, and both electrons are continually dynamically active.

\section{Calculated Momentum Distributions of Ions and Electrons} 

\begin{figure}
\centerline{
	       \includegraphics[width=1.55 in]{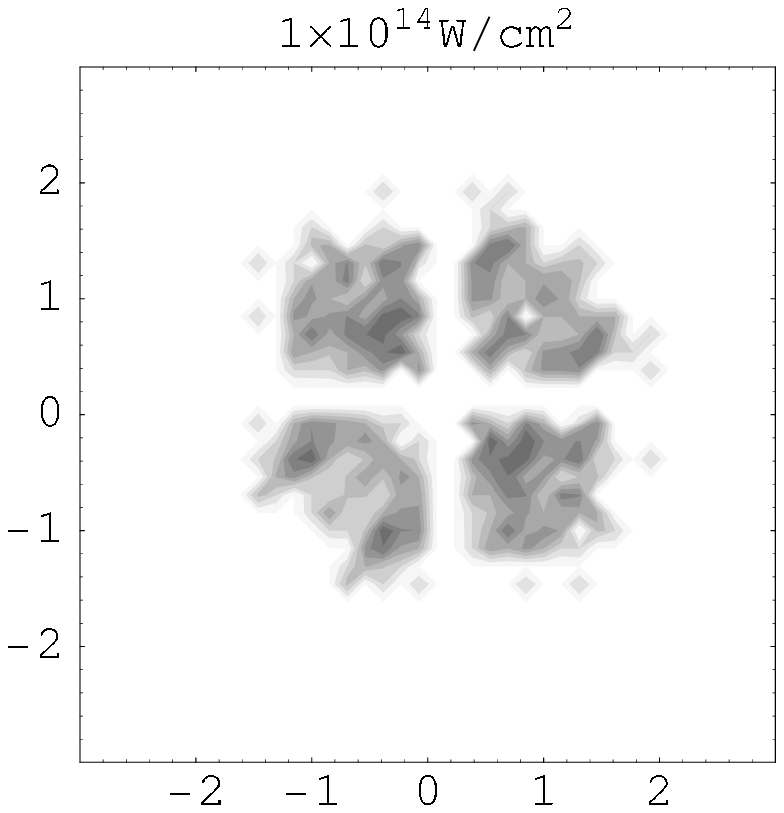}
           \includegraphics[width=1.55 in]{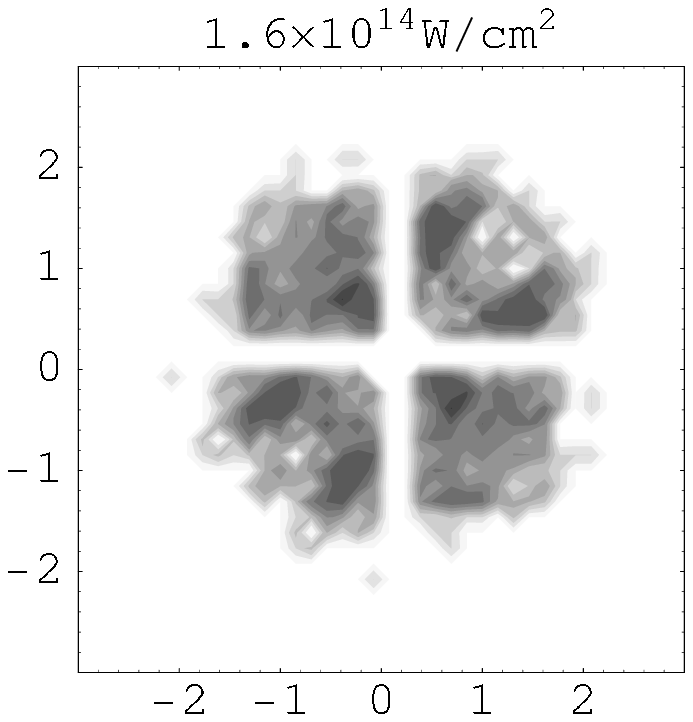}
           }
\centerline{
		   \includegraphics[width=1.55 in]{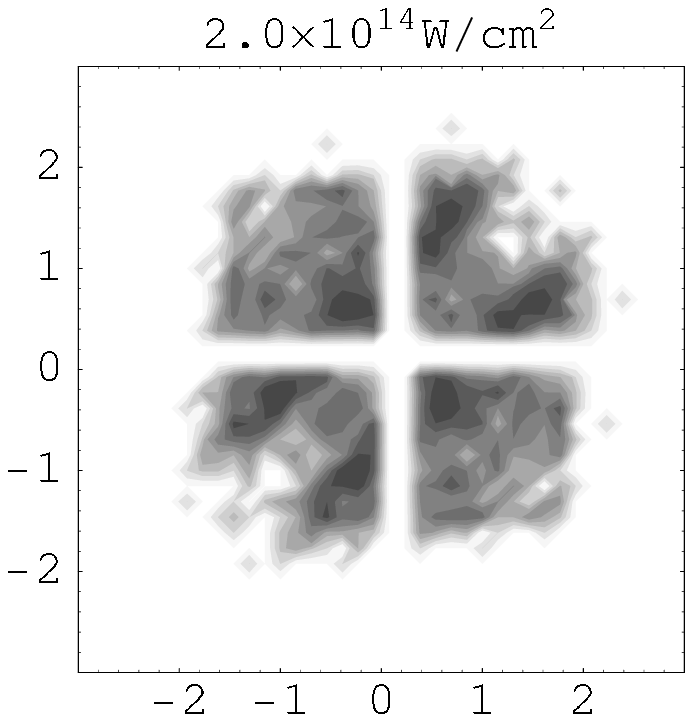}
           \includegraphics[width=1.55 in]{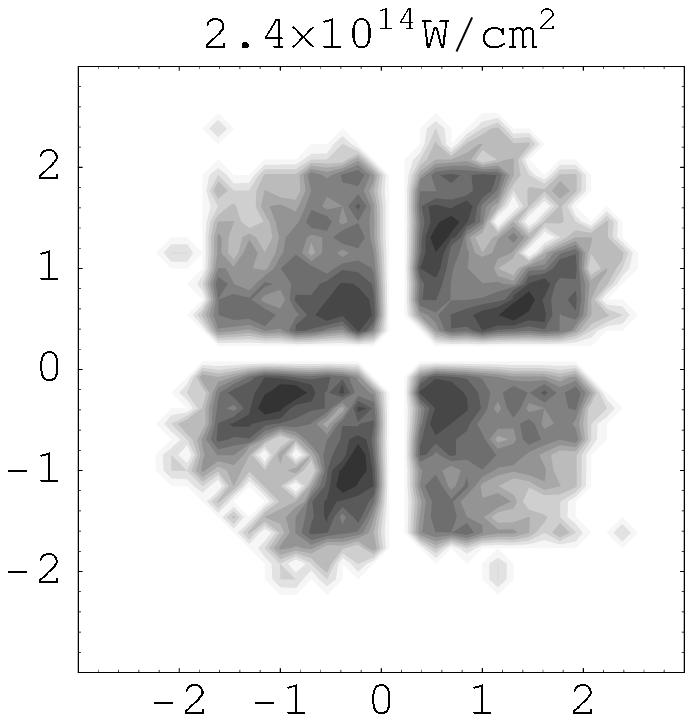}
           }
\centerline{
           \includegraphics[width=1.55 in]{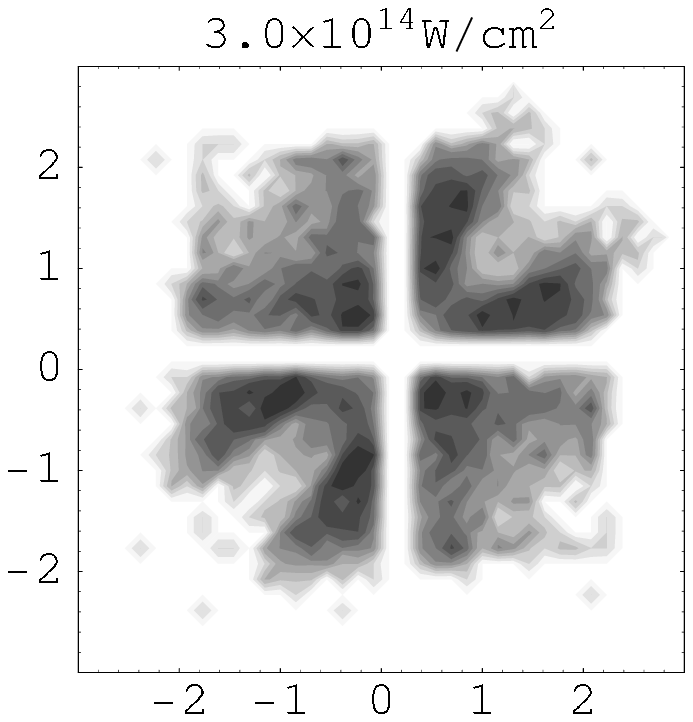}
           \includegraphics[width=1.55 in]{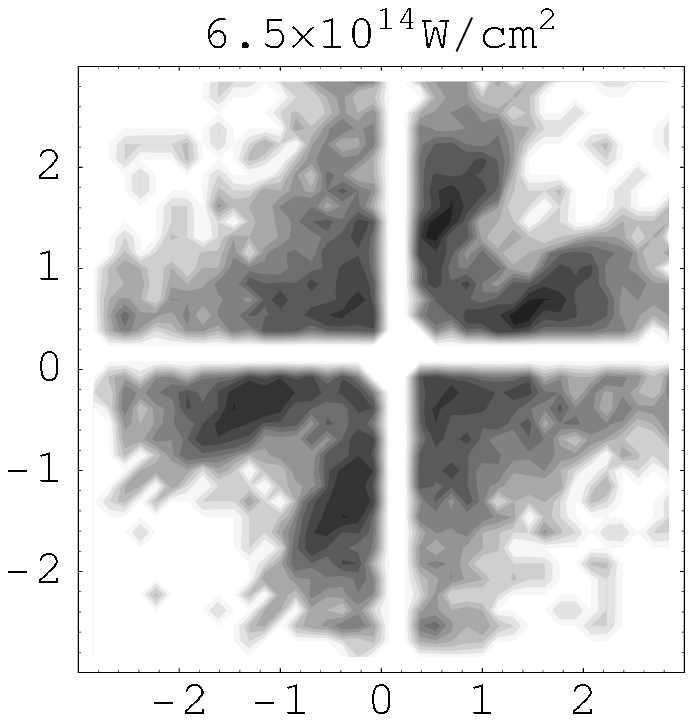}
           }
\caption{End of pulse momentum distributions of doubly ionized electrons at various intensities.  These are density plots in log scale, in which regions with darker color have relatively higher densities of two-electron trajectories. The box size in the figures is much smaller than the computational zone and was chosen for ease of viewing. It does not represent a boundary for the electrons. Notice that there are no trajectories close to the two axes due to our definition of double ionization, in which an electron is considered ionized if its kinetic energy is greater than its nuclear binding potential energy \cite{Panfili-etal02}.   If the definition is modified, the fine details of these distributions will change but the main features will remain the same.}
\label{fig.electronP}
\end{figure}

\begin{figure*}
\centerline{\includegraphics[width=6.6 in]{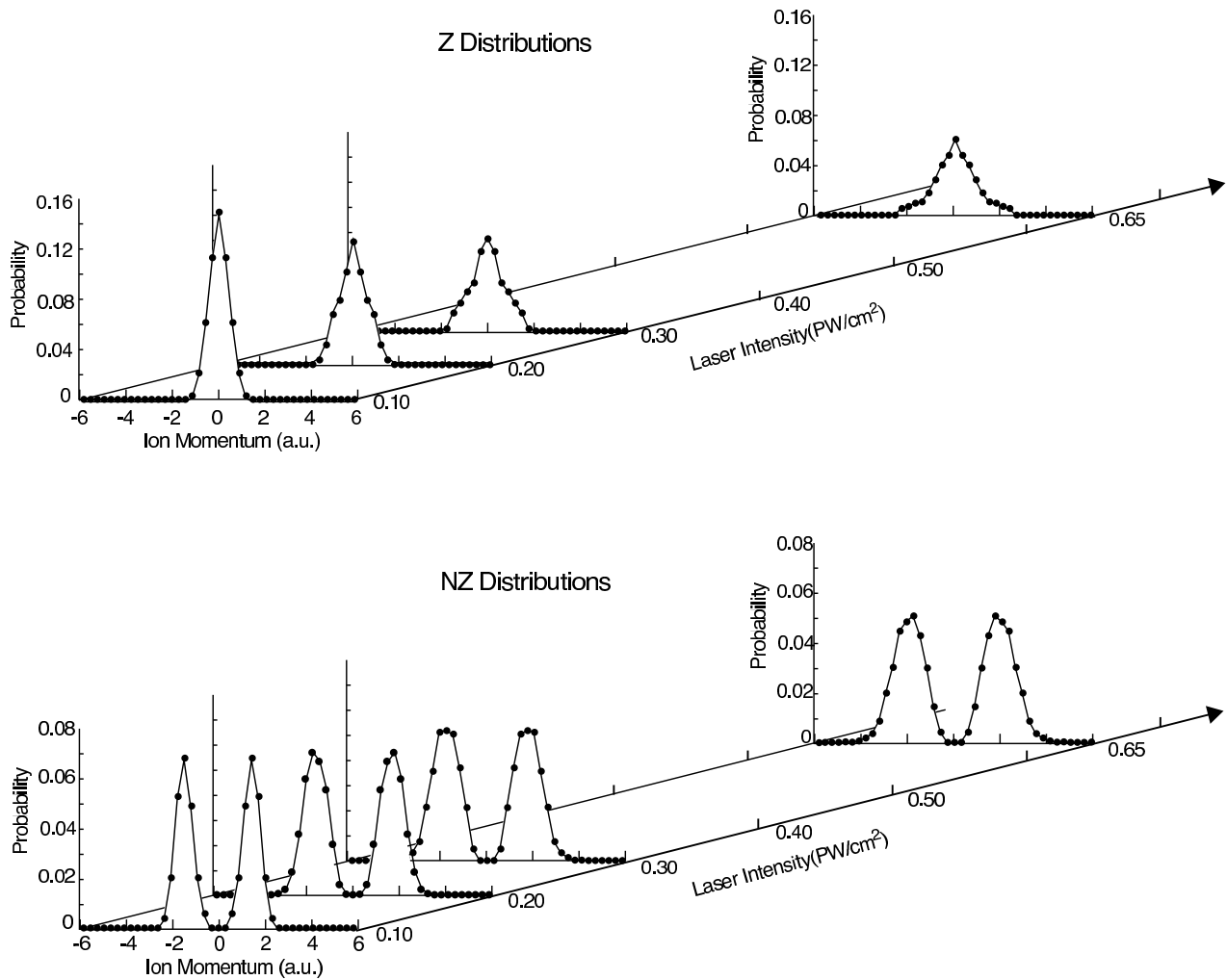}}
\caption{The recoil ion momentum distribution of Z and NZ trajectories at different laser intensities.  Each distribution is calculated by first projecting the two-electron momentum points in Fig. \ref{fig.electronP} onto the diagonal axis.  The raw data projections are then grouped in bins with width 0.3 a.u., and their number is divided by the total number of NSDI points to give the probabilities plotted here.}
\label{fig.ZNZpdist}
\end{figure*}

Calculations based on the ensemble method provide a wealth of ``experimental" information at every intensity. Fig.\ \ref{fig.electronP} shows the two-electron momentum plane at the end of the laser pulse for each of six different laser intensities. The planes contains dense distributions of doubly ionized electron-pair points. Electron pairs with opposite momenta (Z category) produce points in the second and fourth quadrants, and electron pairs moving together (NZ category) provide the points in the first and third quadrants. These points yield theoretical ion momentum distributions for the Z and NZ pairs, and these distributions are found to have different shapes, and also different behaviors as a function of intensity. The calculated shapes are shown with their intensity dependences in Fig. \ref{fig.ZNZpdist}. In both categories the shapes broaden with increasing intensity, more so in the NZ case. More significantly, in both categories the peak of the distribution drops with increasing intensity, but this effect is much stronger for the Z category. Analysis of the time evolution of Z and NZ trajectories, and their relative ability to produce the NSDI ``jets" that have been discussed previously \cite{haan-etal02, Panfili-etal02}, will be reported separately.

\begin{figure}[!ht]
\centerline{\includegraphics[width=2.8 in]{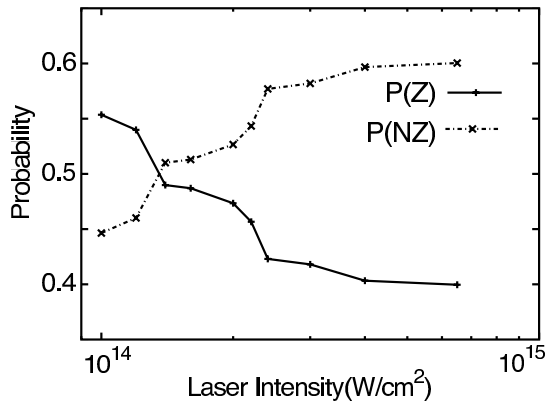}}
\caption{The fractional yield of Z and NZ trajectories at different laser intensities.  Note that P(Z) and P(NZ) indicate the probability of yielding Z and NZ trajectories respectively.}
\label{fig.relZNZprob}
\end{figure}

\begin{figure}
\centerline{
	      \includegraphics[width=3.2 in]{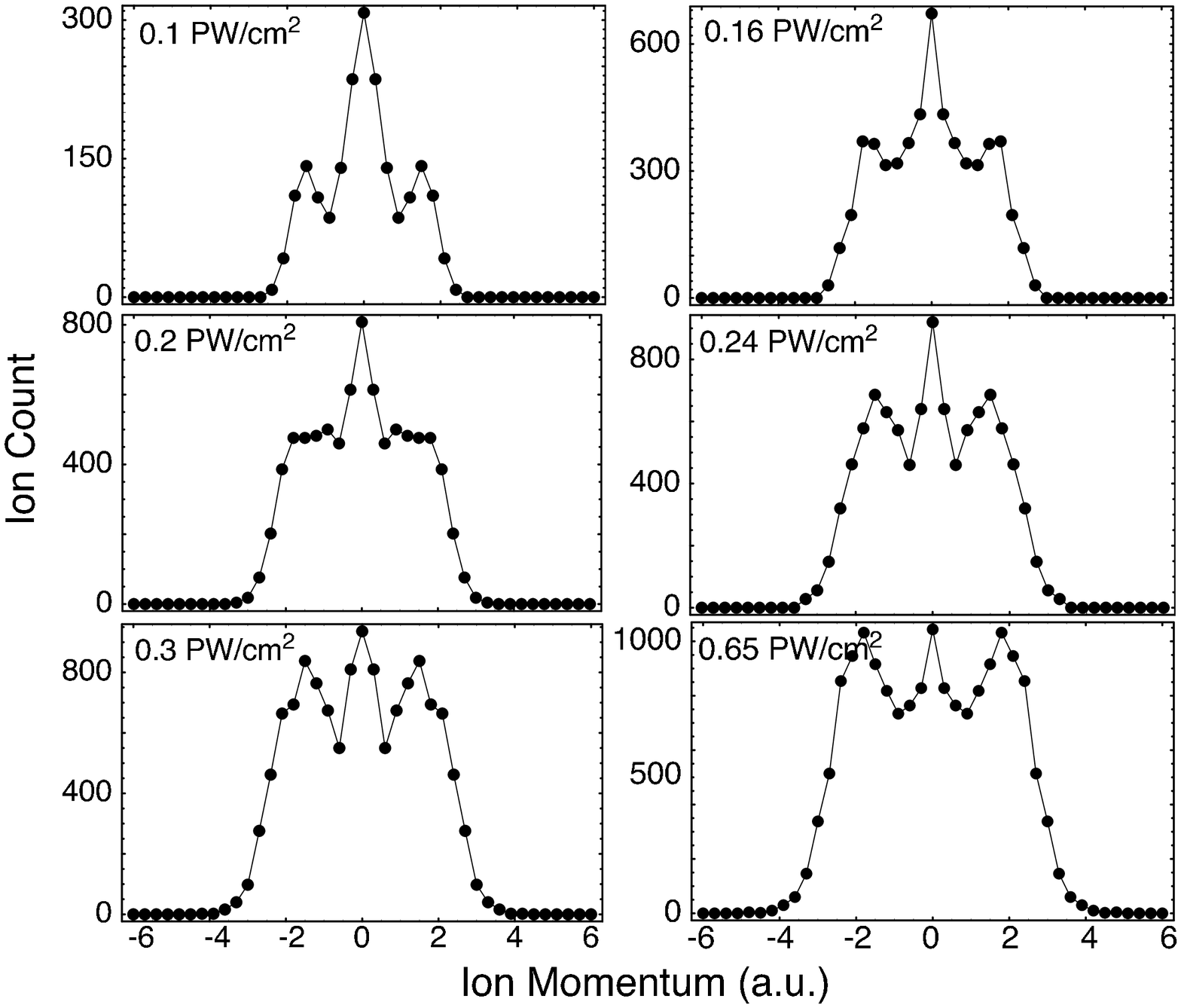}
           }
\caption{Momentum distribution of the recoil ion at various intensities. To reveal the differences at different laser intensity, the calculated data points, which are grouped in bins with width 0.3 a.u., are connected with line.}
\label{fig.recoilp}
\end{figure}

The combined result of broadening and peak-height decrease is a predicted substantial shift in probability in favor of the NZ category. As shown in Fig. \ref{fig.relZNZprob}, the relative probability P(Z) - P(NZ) is predicted to change sign as the intensity exceeds a few times $10^{14}$ watts/cm$^2$, first strongly favoring the Z category at low intensity and then above about $I \sim 5 \times 10^{14}$ watts/cm$^2$ strongly favoring the NZ category. This correlates with the change in appearance of the panels in Fig.\ \ref{fig.electronP} and we can see that this Z $\to$ NZ shift is adequate to explain the shape variation with increasing intensity in the experimental curves shown in Fig. \ref{fig.pdist-expt}. Additional quantitative comparisons can be obtained if the Z and NZ data are combined to make calculated total-momentum distributions.

Fig.\ \ref{fig.recoilp} plots the calculated ion momentum distributions obtained by combining Z and NZ contributions for intensities in the range of $I = 1 - 6.5\times10^{14} W/cm^2$, similar to the range in some of the experiments. Details of the distributions depend to a degree on the definition of double ionization, as explained already in Fig.~\ref{fig.electronP}, but the main characteristics are quite stable. By comparing our calculated results with those measured experimentally (recall the data in Fig. \ref{fig.pdist-expt}), we find the following  similarities in the laser intensity dependence of the shape of the momentum distribution:\\ 
(1) The width of the distribution increases with the laser intensity. The helium experiment shows that the base width of the distribution changes from 8 a.u. at $3.5 \times10^{14} W/cm^2$ to 12 a.u. at $12.5 \times10^{14} W/cm^2$, and the argon experiment shows that the base width of the distribution changes from 10 a.u. at $2.0 \times10^{14} W/cm^2$ to more than 12 a.u. at $3.75 \times10^{14} W/cm^2$.  Similarly, our calculated distribution has the base width of 6 a.u. at  $I=1.0\times10^{14}W/cm^2$.  This width is raised more than 50\% trajectories to about 10 a.u. at $I=6.5\times10^{14}W/cm^2$.\\ 
(2) More low or zero momentum ions are observed at low laser intensities, whereas more high or nonzero momentum ions are observed at high laser intensities.  Such a transition is clearly illustrated both in the experiment and our calculation, in which the side peaks become higher and wider as the laser intensity increases.

One can see from the calculated two-electron momentum distributions that four NZ regions near one axis but relatively far from the origin have the highest density of NZ trajectories, and two Z regions near the origin have the highest density of Z trajectories.  The presence of these densely populated regions indicates that the momenta of two electrons are correlated.  We find that our classical calculations show similarities with the quantum results of Lein, ${\it{et~al.}}$ \cite{lein}, obtained by solving the time-dependent Schr\"odinger equation for the same model with similar parameters. The locations of the highest density two-electron trajectories are in good agreement with the quantum calculations.  This is a further confirmation that the dynamics of two electrons in a strong laser field has definite classical attributes.

\section{Summary}

We have used a completely classical method to investigate the laser intensity dependence of the momentum distributions of electrons and ions.  Based on the final momenta of two electrons, we have identified two distinct categories of electron trajectories, Z and NZ.  We found that the shapes of the ion momentum distributions of the Z and NZ trajectories are very different, in which the Z and NZ trajectories produce single-peak and double-peak structures respectively.  Furthermore, the height and width of their distributions depend on the laser intensity.  Thus, these two types of trajectories act as a basis to give different shapes of the ion momentum distributions at different laser intensities.  Our calculated results showed that the Z trajectories are more favorable at low laser intensities, whereas the NZ trajectories are more favorable at high laser intensities.  This Z $\to$ NZ shift has been demonstrated in the helium and argon experiments.  

Currently, the available experimental ion momentum distributions of neon show a double-peak structure, and its width decreases with decreasing laser intensity within the NSDI regime (see \cite{moshammer_Ne, Ullrich-jpb04, Ullrich-04}).  Since this shape and its variation are consistent with the behavior of the NZ distribution,  we suggest that these neon data are in the intensity regime that favors the production of NZ trajectories \cite{Neon-note}.  By lowering the laser intensity further, our analysis suggests that experiment on neon could observe a sngle-peak momentum distribution that has dominant Z ions.

\section{Acknowledgement}

This work was supported by NSF grant PHY-0072359.  The author wants to thank J. H. Eberly for his suggestions and careful reading of the manuscript.  Also, the author acknowledges the on-going collaboration with S. L. Haan at Calvin College.

\bibliography{apssamp}

\end{document}